\documentclass[journal=jacsat,manuscript=article]{achemso}

\usepackage[version=3]{mhchem} % Formula subscripts using \ce{}
\usepackage{graphicx}
\usepackage{dcolumn}% Align table columns on decimal point
\usepackage{bm}% bold math
\usepackage{siunitx}
\usepackage{soul}
\usepackage{subcaption}
\usepackage{amsmath}
\usepackage{indentfirst}
\usepackage{etoolbox} % needed for patch
\SectionNumbersOn

\author{Jia-Xin Zhu}
\email{jiaxinzhu@stu.xmu.edu.cn}
\affiliation[Xiamen University]
{State Key Laboratory of Physical Chemistry of Solid Surfaces, \textit{i}ChEM, College of Chemistry and Chemical Engineering, Xiamen University, Xiamen 361005, China}

\author{Jun Cheng}
\email{chengjun@xmu.edu.cn}
\affiliation[Xiamen University]
{State Key Laboratory of Physical Chemistry of Solid Surfaces, \textit{i}ChEM, College of Chemistry and Chemical Engineering, Xiamen University, Xiamen 361005, China}
\alsoaffiliation[IKKEM]
{Laboratory of AI for Electrochemistry (AI4EC), IKKEM, Xiamen 361005, China}
\alsoaffiliation[Xiamen University]
{Institute of Artificial Intelligence, Xiamen University, Xiamen 361005, China}

\title
  {Machine Learning Potential for Electrochemical Interfaces with Hybrid Representation of Dielectric Response}

\begin{document}

%%%%%%%%%%%%%%%%%%%%%%%%%%%%%%%%%%%%%%%%%%%%%%%%%%%%%%%%%%%%%%%%%%%%%
%% The "tocentry" environment can be used to create an entry for the
%% graphical table of contents. It is given here as some journals
%% require that it is printed as part of the abstract page. It will
%% be automatically moved as appropriate.
%%%%%%%%%%%%%%%%%%%%%%%%%%%%%%%%%%%%%%%%%%%%%%%%%%%%%%%%%%%%%%%%%%%%%
% \begin{tocentry}

% Some journals require a graphical entry for the Table of Contents.
% This should be laid out ``print ready'' so that the sizing of the
% text is correct.

% Inside the \texttt{tocentry} environment, the font used is Helvetica
% 8\,pt, as required by \emph{Journal of the American Chemical
% Society}.

% The surrounding frame is 9\,cm by 3.5\,cm, which is the maximum
% permitted for  \emph{Journal of the American Chemical Society}
% graphical table of content entries. The box will not resize if the
% content is too big: instead it will overflow the edge of the box.

% This box and the associated title will always be printed on a
% separate page at the end of the document.

% \end{tocentry}

%%%%%%%%%%%%%%%%%%%%%%%%%%%%%%%%%%%%%%%%%%%%%%%%%%%%%%%%%%%%%%%%%%%%%
%% The abstract environment will automatically gobble the contents
%% if an abstract is not used by the target journal.
%%%%%%%%%%%%%%%%%%%%%%%%%%%%%%%%%%%%%%%%%%%%%%%%%%%%%%%%%%%%%%%%%%%%%
\begin{abstract}
  Understanding electrochemical interfaces at a microscopic level is essential for elucidating important electrochemical processes in electrocatalysis, batteries and corrosion. While \textit{ab initio} simulations have provided valuable insights into model systems, the high computational cost limits their use in tackling complex systems of relevance to practical applications. Machine learning potentials offer a solution, but their application in electrochemistry remains challenging due to the difficulty in treating the dielectric response of electronic conductors and insulators simultaneously. In this work, we propose a hybrid framework of machine learning potentials that is capable of simulating metal/electrolyte interfaces by unifying the interfacial dielectric response accounting for local electronic polarisation in electrolytes and non-local charge transfer in metal electrodes. We validate our method by reproducing the bell-shaped differential Helmholtz capacitance at the Pt(111)/electrolyte interface. Furthermore, we apply the machine learning potential to calculate the dielectric profile at the interface, providing new insights into electronic polarisation effects. Our work lays the foundation for atomistic modelling of complex, realistic electrochemical interfaces using machine learning potential at \textit{ab initio} accuracy.
\end{abstract}

% \newpage
% \twocolumn

\section{Introduction}
In modern society, electrochemistry plays an increasingly important role in many areas including material synthesis \cite{zhangElectrochemicalDepositionUniversal2020,leechPracticalGuideElectrosynthesis2022}, energy conversion \cite{norskovOriginOverpotentialOxygen2004,irvineEvolutionElectrochemicalInterface2016}, and energy storage \cite{goodenoughLiIonRechargeableBattery2013}. However, our understanding of the structures and activity of electrochemical interfaces, where the electrochemical reactions happen, is far short of expectation \cite{ofelectrochemistryTopTenScientific2024}. Even for platinum, one of the most important catalysts in electrochemistry \cite{norskovTrendsExchangeCurrent2005,shengCorrelatingHydrogenOxidation2015,feliuPtSingleCrystal2024}, the understanding of how the interfacial structures vary with electrode potentials and electrolyte compositions is still lacking, although it has been widely observed and currently under active investigations that the interfacial structures can be tuned by protons, ions and additives, thus affecting the activity. \cite{leMolecularOriginNegative2020,ojhaDoublelayerStructurePt2022,huangZoomingInnerHelmholtz2023,doblhoff-dierElectricDoubleLayer2023,zhangMeasurementInterpretationDouble2023}. Indeed, the electrochemical characterisations, such as the cyclic voltammograms \cite{rizoUnderstandingInterfacialPH2015,chenCoadsorptionCationsCause2017,chenAdsorptionProcessesPd2020} and the electrochemical impedance spectroscopy \cite{pajkossyImpedanceSpectroscopyInterfaces2005,conwayImpedanceBehaviourProcesses2008}, provide initial but valuable insights into interfacial electrochemistry. Development and application of the in-situ spectroscopic \cite{liSituProbingElectrified2019,liuSituElectrocatalyticInfrared2021,chaoRecentAdvancementsElectrochemical2024} and scanning probe microscopic \cite{grosseDynamicTransformationCubic2021,bianScanningProbeMicroscopy2021} techniques make it possible to investigate the microscopic structures of interfaces. On the other hand, computation and simulation complement experimental measurements in recent decades \cite{norskovOriginOverpotentialOxygen2004,leMolecularOriginNegative2020,grossInitioSimulationsWater2022}, helping interpret the spectra and elucidate the interfacial processes at the atomic level. 

Although atomistic modelling can offer detailed microscopic information on electrochemical interfaces, it often faces a dilemma between accuracy and efficiency. For example, the hydrogen bonding network has been proposed to play an important role in the thermodynamics and the kinetics of the interfacial processes \cite{ledezma-yanezInterfacialWaterReorganization2017, liHydrogenBondNetwork2022, wangEnhancingOxygenReduction2021,maierHowMetalInsulator2024}. While the atomic-level picture of the hydrogen bonding network is difficult to probe in experiment, it can be obtained from molecular dynamics (MD) simulations \cite{shinWaterInterfacialHydrogen2018,liHydrogenBondNetwork2022}. Notably, electronic structures should be included in simulation to describe the potential-dependent hydrogen bonding network accurately in many important systems. For instance, it has been shown that water chemisorption on Pt can induce electron redistribution at the interface and thus have important impacts on interfacial water structures and the differential capacitances of the electric double layer (EDL) \cite{leMolecularOriginNegative2020}. 
\textit{Ab initio} molecular dynamics (AIMD), accounting for both electronic structures and molecular dynamics, can in principle provide accurate descriptions of electrochemical interfaces. However, its high computational cost limits its applications to systems of hundreds of atoms at the time scale of tens of picoseconds, which is insufficient for complex electrochemical interfaces at realistic conditions.  

Fortunately, emerging machine learning (ML) techniques provide us with opportunities to achieve a better compromise between accuracy and efficiency \cite{behlerGeneralizedNeuralnetworkRepresentation2007,bartokGaussianApproximationPotentials2009,batznerEquivariantGraphNeural2022,zengDeePMDkitV2Software2023}. While maintaining the \textit{ab initio} accuracy, the calculations with machine learning potentials (MLPs) are not only $10^3-10^5$ times faster than those with the density functional theory (DFT) but show nearly linear scaling between computational cost and system size \cite{jiaPushingLimitMolecular2020,moAccurateEfficientMolecular2022}. As a result, slow processes in systems with complex structures become accessible by simulations with MLPs \cite{maDynamicCoordinationCations2019,yangSimulatingSegregationTernary2022,gongMachineLearningMolecular2024}. 

Despite the promising progress that has been made, the application of the MLPs in electrochemistry is non-trivial. Currently, the most widely used MLPs are short-ranged due to the finite cutoff distance (i.e., typically smaller than \SI{8}{\angstrom}) used to construct the descriptors \cite{bartokRepresentingChemicalEnvironments2013,zhangEndtoendSymmetryPreserving2018,batznerEquivariantGraphNeural2022}. Such short-range MLPs, as shown in previous work \cite{natarajanNeuralNetworkMolecular2016,jinnouchiFirstprinciplesHydrationFree2021}, are capable of describing the structures at the neutral metal/water interfaces, probably due to the large dielectric constant of the water at the zero electric field limit \cite{fiedlerFullSpectrumHighResolutionModeling2020,zhangNoteDielectricConstant2018}. The lack of explicit long-range interactions in these MLPs, however, has been shown to lead to  breaking of electroneutrality at the bulk electrolyte region when ions are present in the system \cite{zhangElectricalDoubleLayer2024}. Therefore, the short-range MLPs would fail in modelling the electrochemical interfaces under the potential of zero charge (pzc) conditions, let alone the EDL under a potential bias. Strategies to include long-range interactions have been proposed in some ML models in literature \cite{artrithHighdimensionalNeuralnetworkPotentials2011,grisafiIncorporatingLongrangePhysics2019,zhangDeepPotentialModel2022,dufilsPiNNwallHeterogeneousElectrode2023}. One of the commonly used recipes is to separate the total energy into the short- and long-range interactions. While the long-range interaction is approximated with the electrostatic interaction and can be calculated analytically, the remaining short-range interaction can be predicted via the short-range MLPs \cite{artrithHighdimensionalNeuralnetworkPotentials2011,dufilsPiNNwallHeterogeneousElectrode2023, zhangDeepPotentialModel2022}. Instead of separating the short-range and the long-range interactions, it is also possible to include the long-range interaction by embedding the long-range information into the descriptors \cite{grisafiIncorporatingLongrangePhysics2019}.

Although the long-range interactions have been included in some existing ML models, there is another challenge in simulating the electrochemical interface with MLPs, i.e., properly describing the dielectric response of the system. This means that the MLPs should be able to describe the interfacial polarisation at given atomic configurations and (electrostatic) boundary conditions. Recently, many efforts have been made to develop the MLPs with dielectric response \cite{zhangDeepPotentialModel2022,dufilsPiNNwallHeterogeneousElectrode2023,gaoSelfconsistentDeterminationLongrange2022,grisafiPredictingChargeDensity2023b,lewisPredictingElectronicDensity2023a,fallettaUnifiedDifferentiableLearning2024,jollMolecularDynamicsSimulation2024}. For example, some MLPs accounting for electronic polarisation based on local chemical environments\cite{zhangDeepPotentialModel2022,gaoSelfconsistentDeterminationLongrange2022,lewisPredictingElectronicDensity2023a,fallettaUnifiedDifferentiableLearning2024,jollMolecularDynamicsSimulation2024}, are suitable for describing the dielectric response of insulators, but fail in treating the non-local charge transfer in conductors due to the short-sightedness of the descriptors. In order to overcome the shortcoming of short-sightedness, other forms of MLPs have been proposed on the basis of the charge equilibrium (QEq) scheme \cite{ghasemiInteratomicPotentialsIonic2015,dufilsPiNNwallHeterogeneousElectrode2023,koAccurateFourthGenerationMachine2023,dufilsPiNNwallHeterogeneousElectrode2023}, allowing for the non-local charge transfer. However, these MLPs inherit the intrinsic limitations of the QEq method originating from the semi-local approximations of kinetic energies. As a consequence, not only would the QEq-based model overestimate the polarisability of insulators, but also incorrectly predict fractional rather than integer charged fragments when a molecule dissociates, which limits its applicability to isolated molecules or clusters \cite{verstraelenACKS2AtomcondensedKohnSham2013,verstraelenCanElectronegativityEqualization2015,shaoFinitefieldCouplingLearning2022}.

In this work, we propose a hybrid scheme of MLPs to treat electrochemical interfaces (ec-MLP), which integrates the dielectric responses of electronically insulating electrolytes and conducting metal electrodes. In particular, the dielectric response of electrolytes can be obtained by the shift of the positions of the Wannier centroids, which is defined as the average positions of the maximally localised Wannier centres (MLWCs) and predicted via a Deep Wannier (DW) deep neural network model based on the local chemical environment\cite{zhangDeepPotentialModel2022}. Meanwhile, the dielectric response of the metal electrode can be taken into account via the Siepmann–Sprik model\cite{siepmannInfluenceSurfaceTopology1995}. With this method, the atomic charges of the electrode can be solved by minimising the electrostatic energy under the given atomic configurations and the boundary condition. Knowing the charge distributions of both electrolytes and electrodes, the long-range electrostatic energy can be calculated analytically and extracted from the total energy. The residual energy term, which is regarded to be dependent on the local chemical environment, is described via a short-range ML model. 
The feasibility of the ec-MLP proposed in this work is illustrated in an important and challenging model system, i.e., the Pt(111)/KF(aq) interfaces. For this model system, the ec-MLP can describe not only the water chemisorption phenomenon but also the potential-dependent interfacial structures and differential capacitances. Based on the well-validated ec-MLP, we further calculate the dielectric profile at the Pt/water interface, which reveals the importance of electronic effects and offers new insights into the molecular origins of interfacial dielectric properties. Overall, the ec-MLP proposed in this work provides access to going beyond the timescale and the length scale of \textit{ab initio} simulation, showing great potential in future electrochemical studies.

\section{Results}
\subsection{Machine learning model with hybrid representation of dielectric response at metal electrode/electrolyte interfaces}

At electrochemical interfaces, there can exist an extremely large electric field\cite{staffaDeterminationLocalElectric2017,schwarzElectrochemicalInterfaceFirstprinciples2020}, the influence of which should be considered when simulating the electrochemical interface. The first challenge to be overcome is to describe how the interfacial polarisation varies with the electric field, i.e., the dielectric response of the ionically conducting electrolyte and electronically conducting electrode at the interface. In the theory of dielectrics, the species in the dielectrics responds to the electric field in the following ways, i.e., the translation of ions, the reorientation of solvent molecules, the vibration of bonds and the polarisation of electrons \cite{schwarzElectrochemicalInterfaceFirstprinciples2020}. Concerning the electrolyte, which is an ionic conductor and electronic insulator, the first three terms predominate in the dielectric response. The metal electrode, on the contrary, is the electronic conductor and responds to the electric field by electronic polarisation. The distinct dielectric response of electrolyte and electrode leads to different treatments when describing the response in the ec-MLP, i.e., the Wannier centroid method for electrolytes and the polarsiable electrode method for electrodes.

In the electronically insulating electrolyte, the dielectric response is mainly undertaken by the motion of atoms and ions. We therefore choose the WC method to describe this dielectric response. In this method, the positions of WCs relative to the associated nuclei are predicted based on the local chemical environment \cite{zhangDeepPotentialModel2022}. 
% When applying an electric field, the charged particles move in the field, and the dielectric response appears. 
% It is notable that while the electronic polarisation induced by the change in the local chemical environment has been considered, that induced by the external electric field is omitted currently. Nevertheless, it is still a reasonable approximation, especially around the pzc, which will be justified qualitatively in the next section. 
% This response, in principle, can be described by assigning proper atomic charges. When applying an electric field, the charged particles move in the field, and the dielectric response appears. 
% In order to capture this dielectric response from the motion of atoms and ions, we describe the charge distribution of the electrolyte with the point charges located at the nuclei and the WCs. In the modern theory of polarisation \cite{Marzari1997,restaElectronLocalizationInsulating1999,Spaldin2012}, the MLWCs can be regarded as the equivalent positions of electron pairs, and the displacements of the MLWCs in the internal coordinates indicate the polarisation of electrons. Usually, the MLWCs can be uniquely assigned to specific central atoms. As a result, it is convenient to define the average positions of the MLWCs as the WCs if we are interested in the dielectric response. 
% However, it might lead to an issue when the atoms are ``frozen'' at high surface charge densities. Therefore, the accuracy of the model should be validated qualitatively under the electrochemical conditions of interest, as will be shown in the next section.
In contrast to the electrolyte, the motion of the atoms is usually neglectable and the dielectric response from the electronic polarisation (i.e., the electronic dielectric response) plays an important role in the metal electrode. Previously, there have been some methods developed for describing the electronic dielectric response of metal (i.e., polarisable electrode method) \cite{geadaInsightInducedCharges2018,siepmannInfluenceSurfaceTopology1995,nakanoChemicalPotentialEqualization2019}. For example, in the Siepmann–Sprik model\cite{siepmannInfluenceSurfaceTopology1995}, the charge distribution in the electrode is approximated with the spherical Gaussian charges located at the atom sites. This leads to a quadratic form of the electrostatic energy with respect to the magnitude of the Gaussian charges. The charges obtained by minimising the electrostatic energy vary with the atomic configuration and the boundary condition, accounting for the electronic dielectric response of the electrode. Remarkably, the Siepmann–Sprik model has been used in conjunction with DFT in a hybrid quantum mechanics/molecular mechanics (QM/MM) calculation \cite{golzeSimulationAdsorptionProcesses2013}. When this QM/MM scheme is used to investigate the molecular adsorption on the metal slab, it can not only capture the correct long-range interactions but reproduce the adsorption energies in the corresponding full DFT calculation. The promising result indicates that the Siepmann–Sprik model is a good choice for describing the electronic dielectric response of the metal electrode. 

\begin{figure}
  \centering
  \includegraphics{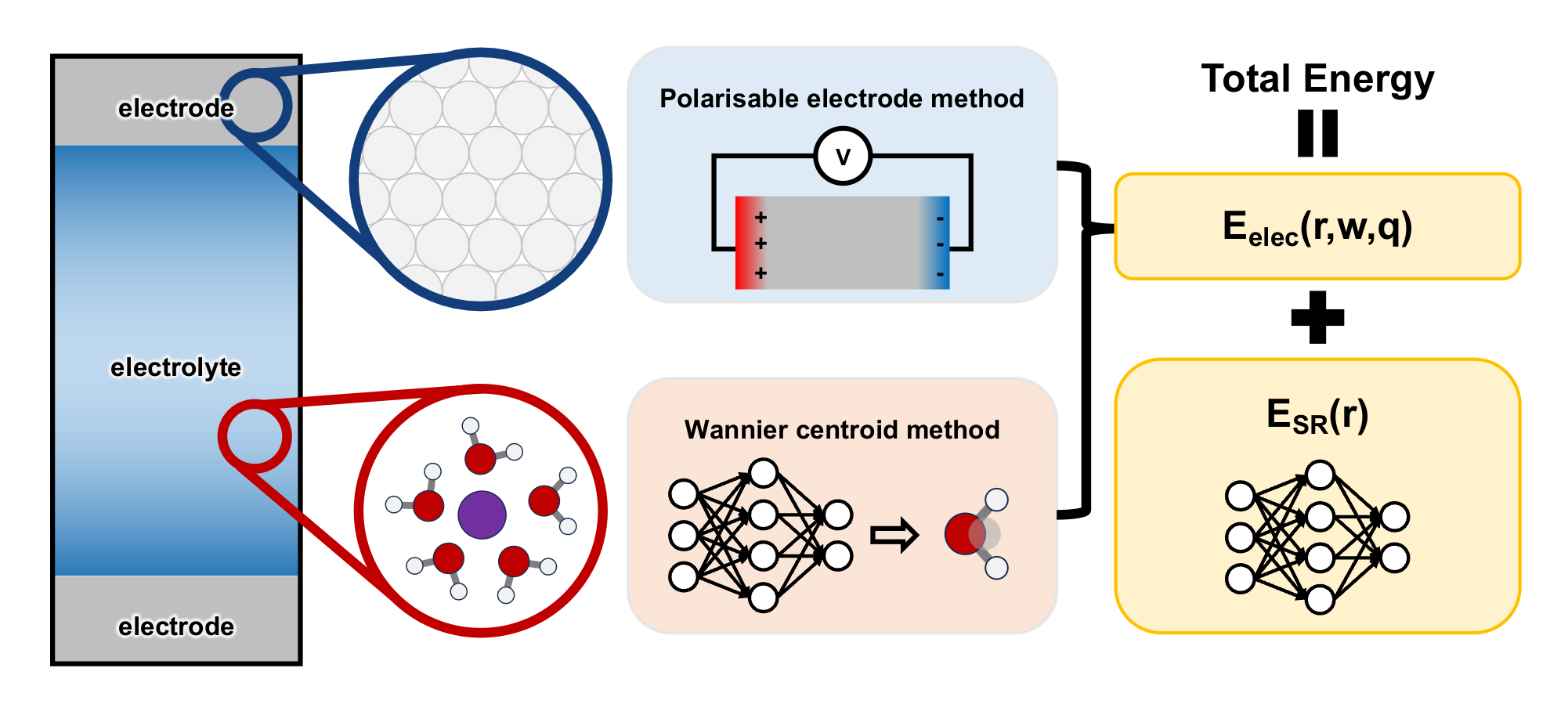}
  \caption{\textbf{Schematic illustration of the machine learning model for electrochemistry (ec-MLP)} The architecture of the ec-MLP is illustrated. The charge distributions in the electrode (i.e., atomic charges $q$) can be predicted with the polarisable electrode method, while those in the electrolyte (i.e., the positions of Wannier centroids $w$ and the positions of nuclei $r$) can be predicted with the ML-based Wannier centroids method. The charge distribution can be used to calculate the long-range electrostatic contribution $E_\text{elec}$ in the total energy and the residual short-range energy $E_\text{SR}$ is predicted by the MLP based on the local chemical environment.}
  \label{fig:model}
\end{figure}

The WC method and the polarisable electrode method can be integrated for a hybrid representation of the dielectric response of interfaces, which is used for the further derivation of total energies in the ec-MLP (see Fig.~\ref{fig:model}). For a given atomic configuration, the positions of WCs are predicted from a ML model for atomic tensorial properties\cite{grisafiSymmetryAdaptedMachineLearning2018,zhangDeepNeuralNetwork2020}, and the charge distribution in electrolytes is approximated as the point charges located at the positions of nuclei and WCs. Based on this charge distribution, the electrostatic potential generated by the electrolyte at the sites of electrode atoms can be calculated, with which the atomic charges of the electrode atoms can be solved at a given electrostatic boundary condition. The charge distribution at the interface (i.e., for both electrolytes and electrodes) allows the calculation of the long-range electrostatic contribution $E_\text{elec}(r,w,q)$ in the total energy $E(r,w,q)$, and the residual short-range energy $E_\text{SR}(r)$ is predicted via a MLP model based on the local chemical environment\cite{artrithHighdimensionalNeuralnetworkPotentials2011,dufilsPiNNwallHeterogeneousElectrode2023, zhangDeepPotentialModel2022}:
\begin{align}
  &E(r,w,q)=E_\text{SR}(r)+E_\text{elec}(r,w,q). \label{eq:ener}
  % &F_I = -\frac{\partial E_{SR}}{\partial r_I}-\frac{\partial E_{elec}}{\partial r_I} - \frac{\partial E_{elec}}{\partial w_{n(I)}} + \sum_n\frac{\partial E_{elec}}{\partial w_n}\frac{\partial w_n}{\partial r_I}, \label{eq:elec_force}
\end{align}
$r$ and $w$ are the positions of the nuclei and the WCs, respectively. $q$ is the magnitude of the charges located at the sites of the nuclei or WCs. In this work, $E_\text{elec}$ in Eq.~\ref{eq:ener} is defined as the reciprocal term in the Ewald summation algorithm. The Deep Wannier (DW) deep neural network model \cite{zhangDeepNeuralNetwork2020} and the Siepmann–Sprik model \cite{siepmannInfluenceSurfaceTopology1995} are chosen to describe the dielectric response in electrolytes and metal electrodes, respectively, in the interest of the practical implementation. Nevertheless, other similar methods can be used instead. For example, the atomic charges projected from the MLWCs can be predicted via a ML model to capture the dielectric response of electrolyte, while the charge equilibrium method\cite{nakanoChemicalPotentialEqualization2019} can be regarded as a generalised form of the Siepmann-Sprik model to describe the dielectric response of electrode.

\subsection{Validations of machine learning potentials for electrochemistry}

In the following, the Pt(111)/KF(aq) interface is chosen as the model system to demonstrate the validity of the ec-MLP proposed above. This choice is motivated by the widespread utilisation of the Pt(111)/electrolyte interface as the model system in both experimental and computational electrochemistry. 
Most importantly, it is well known in the literature that Pt shows a bell-shaped Helmholtz differential capacitance curve \cite{pajkossyOriginDoubleLayer2003} caused by dynamic chemisorption of water induced by the applied potential \cite{leMolecularOriginNegative2020}. This phenomenon on Pt highlights the simultaneous importance of electronic structure and molecular dynamics effects, which cannot be accounted for by static DFT or classical molecular dynamics calculation. Therefore, it serves as an excellent model system for validating ec-MLP in comparison with AIMD.
% Most importantly, electron redistribution induced by water chemisorption happens at Pt/electrolyte interfaces, which cannot be tackled by the classical force field. In other words, successfully simulating the electrified Pt/electrolyte interfaces is a firm validation for the ec-MLP. In this section, we aim to demonstrate the effectiveness of the ec-MLP in accurately describing not only the water chemisorption phenomenon under the pzc but also the potential-dependent changes in the interfacial water structures and the Helmholtz differential capacitance. 

The accuracy of the ec-MLP is first evaluated by comparing the predicted energies, atomic forces and positions of WCs from the MLP with the corresponding values obtained from DFT calculations. A concurrent learning workflow is utilised to explore the configuration space and collect representative configurations for training the ec-MLP\cite{zhangDPGENConcurrentLearning2020}. The details of the training of the ec-MLP and the DFT calculations are provided in the Methods section. As listed in Table~\ref{tab:validation_dft}, the root mean square errors (RMSEs) in the Wannier centroids are \SI{1.9}{m\angstrom} for the training dataset and \SI{2.0}{m\angstrom} for the testing dataset, which are comparable to the uncertainty reported in previous work \cite{zhangDeepPotentialModel2022}. 
% Notably, there are some outliers in the training dataset, which are attributed to some configurations collected under high-temperature conditions (e.g., water dissociation). These configurations are hardly visited in the simulations under the room temperature (e.g., the testing dataset), which are the cases of our interest. Therefore, we justify the minor influence of the poor descriptions of these configurations. 
Additionally, the potential energy surface of the system can be predicted accurately by the MLP. When applied to the training dataset, the ec-MLP yields a RMSE of \SI{1.305}{meV/atom} for total energies and \SI{75.00}{meV/\angstrom} for atomic forces. The testing dataset demonstrates even smaller RMSEs for both energies and atomic forces, highlighting the generalisability of the MLP. 
% It is worth noting that the majority of the dataset consists of data without an external electric field (see Supplementary Sec.1). Despite this, ec-MLP showcases remarkable performance, even in the presence of an external electric field.

\begin{table}
    \centering
    \caption{Root mean square errors (RMSEs) in the positions of WCs (m\si{\angstrom}), the potential energies (\si{meV/atom}) and the atomic forces (\si{meV/\angstrom}) by comparing the results from the MLP and DFT calculation, in either the training set or the testing set. The details of the dataset are referred to Supplementary Sec.1.}
    \begin{tabular}{c|c|c|c}
        \hline
        \hline
         & WC positions & energies & forces \\
        \hline
        training set & 1.9 & 1.305 & 75.00 \\
        testing set & 2.0 & 0.9638 & 57.19 \\
        \hline
        \hline
    \end{tabular}
    \label{tab:validation_dft}
\end{table}

\begin{figure}
  \centering
  \includegraphics{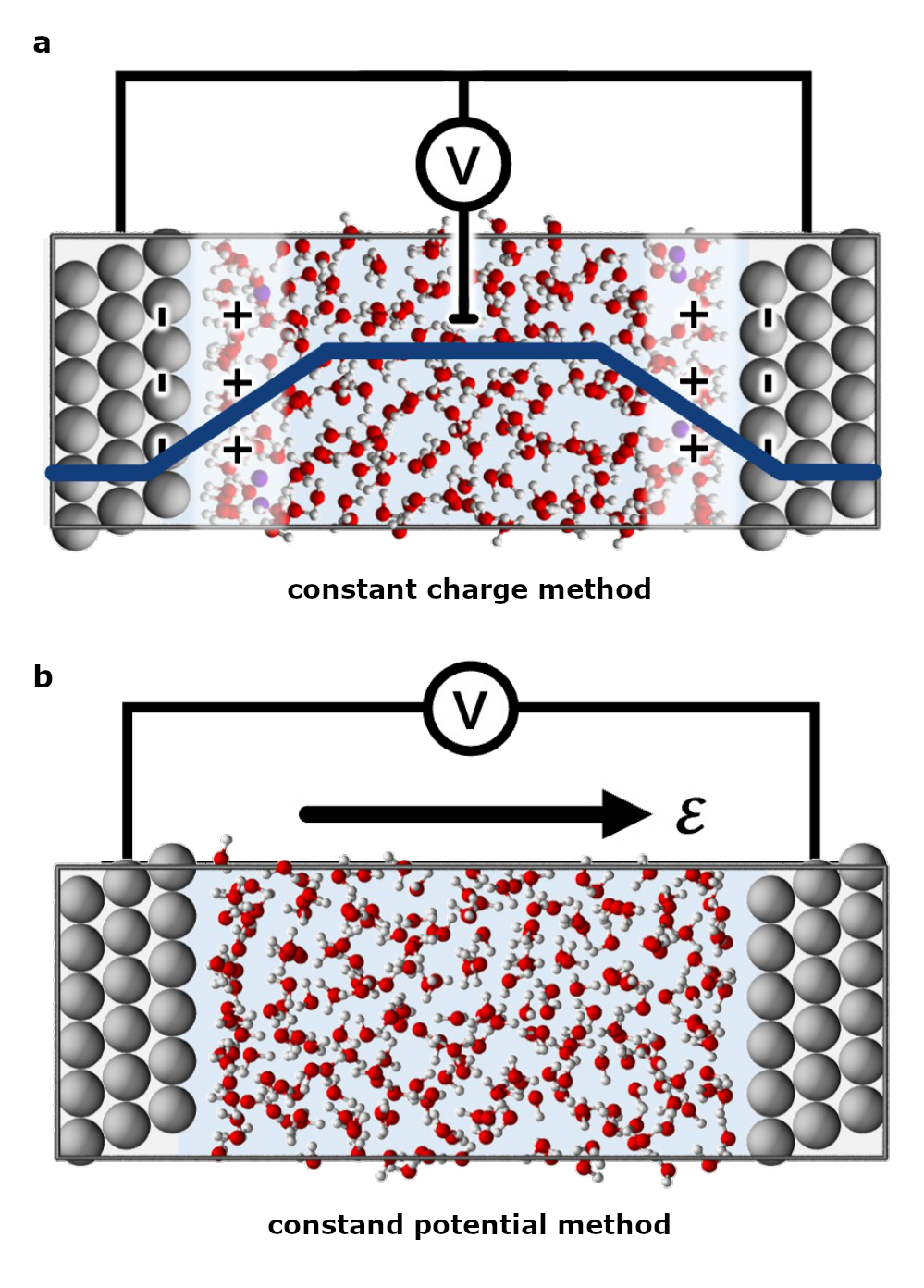}
  \caption{\textbf{Representative snapshots of electrochemical interfaces with different electrifying methods.} \textbf{a} The electrochemical interfaces can be electrified by adding counterions adjacent to electrodes in a symmetric supercell. The electrodes carry the total charges in the opposite sign but the same magnitude as the electrolytes.\cite{leRecentProgressInitio2021} In this case, the electrostatic potential distributions at two interfaces are identical at the thermodynamics limit, as shown by the blue line. \textbf{b} Additionally, it is also feasible to build up a model with both positively and negatively charged interfaces by applying a potential drop (or a periodic electric field $\mathcal{E}$) crossing the supercell (e.g., finite field methods)\cite{nunesBerryphaseTreatmentHomogeneous2001,umariInitioMolecularDynamics2002,souzaFirstPrinciplesApproachInsulators2002,stengelElectricDisplacementFundamental2009}. In all snapshots, the Pt, K, O, and H atoms are coloured in silver, violet, red, and white, respectively. The Pt electrode regions are shaded in grey, while the electrolyte region is shaded in blue.}
  \label{fig:md_model}
\end{figure}

Good performance of the ec-MLP in predicting energies, atomic forces and positions of WCs lays a solid foundation for reproducing the potential-dependent interfacial water structures. Here, we perform MD simulations on the atomic models as shown in Fig.~\ref{fig:md_model}a. The Pt electrodes are described with a (6$\times$6) 6-atomic-layer slab while the $\sim$\SI{2.2}{mol/L} KF aqueous solution is filled between the space of \SI{30}{\angstrom} between two Pt surfaces. To electrify the interfaces, we adopt a similar method, i.e., counterions in symmetric supercells, as in the previous work \cite{leMolecularOriginNegative2020,wangUnderstandingEffectsElectrode2023,leRecentProgressInitio2021}.
In this way, the electrodes carry the total charges in the opposite sign but the same magnitude as the electrolytes, which can be regarded as a reasonable approximation of a compact EDL at the high concentration limit. In the constant charge method, the total charges of electrodes should be defined by users in the ec-MLP, while those in AIMD are induced automatically by the electroneutrality constraint. The electrostatic potential profile is calculated based on the charge distribution by solving the Poisson equation. The timescales for all MD simulations are set as \SI{100}{ps}, which are longer than those used in the previous AIMD simulations (i.e., 10$\sim$\SI{20}{ps}) and lead to better statistical accuracy. The details of the setup for MD simulations are referred to the Methods section. 

\begin{figure}
  \centering
  \includegraphics{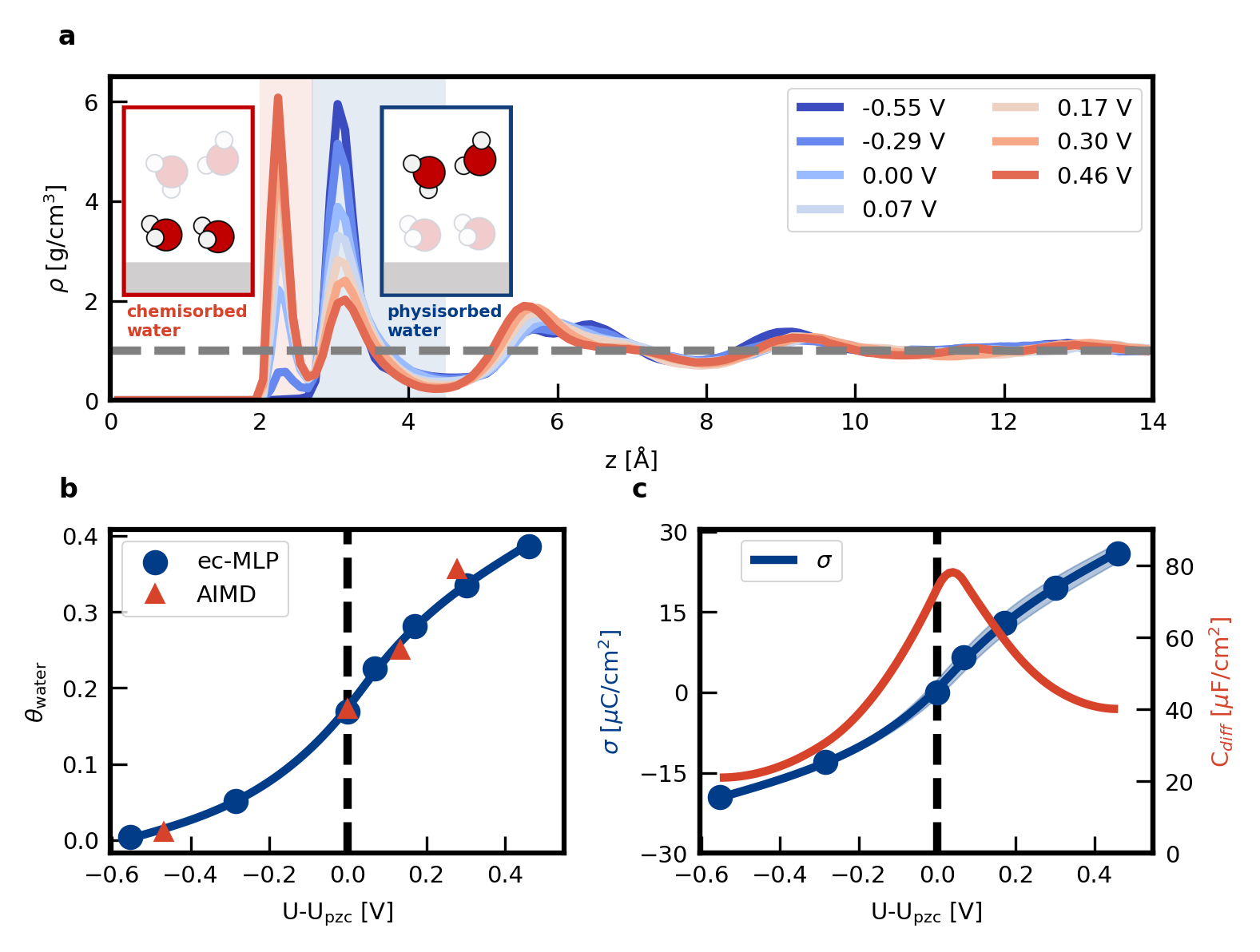}
  \caption{\textbf{Structure and capacitance of the Pt/KF interface calculated using the ec-MLP.} \textbf{a} Water density distribution profile in the direction perpendicular to the Pt surface under different electrode potentials versus pzc. The zero point in the z coordination is set at the average position of the outermost Pt atoms. The chemisorbed water (red) and the physisorbed water (blue) are identified with the shadows in the background, the example configurations of which are shown in the insets. The grey dashed line refers to 1 g/cm$^3$ as a guide to the eye. \textbf{b} Potential-dependent coverage of the chemisorbed water. The blue curve and the red triangles represent the results from the simulations with the ec-MLP in this work and from the AIMD simulations in the previous work \cite{leMolecularOriginNegative2020}, respectively. \textbf{c} Plots of surface charge density $\sigma$ and the differential capacitance as a function of the potential versus pzc $(U-U_\text{pzc})$. The blue curve is the charge density-potential function calculated from the simulations with the ec-MLP, with the uncertainty shown in the blue shadow. The differential capacitance calculated from the first derivative of the blue curve is shown in the red curve.}
  \label{fig:validation_simulation}
\end{figure}

In the simulations under the pzc, the chemisorbed water can be observed in the water density distribution profile in the direction perpendicular to the surface, corresponding to a peak at $\sim$\SI{2.3}{\angstrom} away from the Pt surface (see the insert with red frame in Fig.~\ref{fig:validation_simulation}a). 
% This chemisorbed water adopts an optimal configuration, in which the molecular planes of the water are almost parallel to the surface and the water prefers to form hydrogen bonds with the neighbouring chemisorbed water. 
Furthermore, the potential-dependent behaviours of the water chemisorption can also be captured. When the potential varies from negative to positive values, the coverage of the chemisorbed water increases, which agrees with the previous AIMD simulations (see the blue curve and the red triangles in Fig.~\ref{fig:validation_simulation}b). Meanwhile, the molecular orientation of the chemisorbed water remains almost unchanged. In addition to water chemisorption, water reorientation at the interface can also be described accurately. The water molecules located at 2.7-\SI{4.5}{\angstrom} away from the Pt surface, identified as the physisorbed water, reorientate when changing the electrode potential, leading to a total dipole moment aligning with the electric field. Further details about the potential-dependent orientation of the interfacial water can be found in Supplementary Sec.2. All the results mentioned above, generated from the simulations with the ec-MLP, agree with those from the previous AIMD simulations not only in the qualitative trend but also in the quantitative values \cite{leDeterminingPotentialsZero2017, leStructureMetalwaterInterface2018, leMolecularOriginNegative2020}.

The accurate description of the potential-dependent water chemisorption and reorientation at the interface makes it possible to reproduce the Helmholtz differential capacitance of the EDL. The differential capacitance is defined as the first derivative of surface charge with respect to the potential. In our model, the surface charge is the controlled variable in simulation, which is equal to the number of counterions. The potential can be calculated by referring the average electrostatic potential in the bulk electrode region to that in the bulk electrolyte region \cite{leMolecularOriginNegative2020,leDeterminingPotentialsZero2017}. 
% In this way, the calculated potential differs from the electrode potential relative to a given reference electrode (e.g., standard hydrogen electrode) only by a constant.
As shown in Fig.~\ref{fig:validation_simulation}c, the surface charge density-potential function shows the uncertainty smaller than \SI{0.05}{V}, guaranteeing the statistical accuracy of the differential capacitance curve.

% For a quantitative analysis, we adopt the theoretical model proposed by Le \textit{et al.} to fit the data from our MD simulations. In this model, the coverage of the chemisorbed water is the function of the surface charge density, following the Frumkin isotherm. The water chemisorption induces electronic polarisation at the interface, which influences the interfacial potential difference and hence the capacitance at a given surface charge density. 
Taking the first derivative of the surface charge density-potential function, the differential capacitance can be calculated with its maximum value of $\sim$80 $\mu$F/cm$^2$ around the pzc. In the potential windows chosen in this work, the differential capacitance curve converges to values of 20 and 40 $\mu$F/cm$^2$ at the negative and positive limits, respectively. Overall, the ec-MLP is capable of reproducing the bell shape of the differential capacitance curve, and the capacitance value is in good agreement with the experimental results \cite{pajkossyOriginDoubleLayer2003} and the AIMD results \cite{leMolecularOriginNegative2020, wangUnderstandingEffectsElectrode2023}. Since the bell-shaped capacitance curve can only be captured when taking both electronic structure and molecular dynamics effect into account, this stringent validation strongly shows the suitability and promise of the ec-MLP in simulating electrochemical systems.

\subsection{Dielectric profile of interfacial water}

Water in contact with solids shows distinct dielectric behaviours from bulk water, attracting great interest in both experiment and simulation. For instance, water between two hydrophobic surfaces separated by a few nanometers shows an anomalously low dielectric constant due to strong confinement effects\cite{fumagalliAnomalouslyLowDielectric2018}. At highly heterogeneous electrochemical interfaces, the dielectric constant of water is anticipated to be a complex function of position, which is relevant when analysing the electrochemical conditions for interfacial processes. However, our knowledge about how the dielectric constant varies at the interface is still lacking at the qualitative level. Recently, MD simulations provide access to the dielectric profile, i.e., the dielectric constant distribution in the direction perpendicular to the surface \cite{bonthuisDielectricProfileInterfacial2011,bonthuisProfileStaticPermittivity2012,deissenbeckDielectricPropertiesNanoconfined2021,olivieriConfinedWaterDielectric2021,locheEffectsSurfaceRigidity2022,tranNegativeDielectricConstant2023,deissenbeckDielectricPropertiesNanoconfined2023}. The dielectric profile can be further used in the continuum modelling and lead to a better agreement between the experimental data and the modelling results\cite{schwarzElectrochemicalInterfaceFirstprinciples2020,ringeImplicitSolvationMethods2022}. Nevertheless, the existing studies of dielectric profiles are limited to the cases of inert interfaces, and the calculation of interfaces of high activity (e.g., Pt/water) is still lacking, probably due to the following reasons. On the one hand, simulations with classical force fields cannot describe the interfacial water structures correctly due to the omission of the electronic structure, and the timescale for statistically reliable results is not affordable for AIMD simulations. On the other hand, it is not yet practical to apply an electric field between two metal slabs in current DFT implementations. Fortunately, with the help of the ec-MLP, we can overcome the limitations mentioned above and calculate the dielectric profile at the Pt(111)/water interface for the first time. Furthermore, based on the calculated profile, the impact of the electronic polarisation on the interfacial dielectric response will be revealed.

Aiming for the dielectric profile of water at the Pt/water interface, a small perturbing electric field is applied to the model shown in Fig.~\ref{fig:md_model}b. The local electric fields induced by the applied field can be obtained by integrating the induced charge densities. Consequently, the dielectric profile can be calculated as follows\cite{bonthuisDielectricProfileInterfacial2011}:
\begin{align}
  \varepsilon_\perp^{-1}(z) = \frac{\mathcal{E}_\perp^\text{ind}(z)}{\mathcal{E}_\perp^\text{ind}(z)+\varepsilon_0^{-1}P_\perp^\text{ind}(z)}, \label{eq:inveps}
\end{align}
where $z$ is the distance away from the surface, $\varepsilon_0$ is the vacuum permittivity, and $\varepsilon$ is the dielectric constant (i.e., relative permittivity). $\mathcal{E}_\perp^\text{ind}$ and $P^\text{ind}$ are the induced electric field and polarisation, respectively. The footnote ``$\perp$'' refers to the direction perpendicular to the surface. Details of calculating $\mathcal{E}_\perp^\text{ind}$ and $\varepsilon_0^{-1}P_\perp^\text{ind}$ are referred to Supplementary Sec.3. Notably, a large fluctuation can be observed in the resulting dielectric profile, which includes the effects of the intramolecular variations in the electric field. Since it is only meaningful to consider the response on length scales corresponding to intermolecular distances, the dielectric profile should be averaged over the regions of interest. 

\begin{figure}
  \centering
  \includegraphics{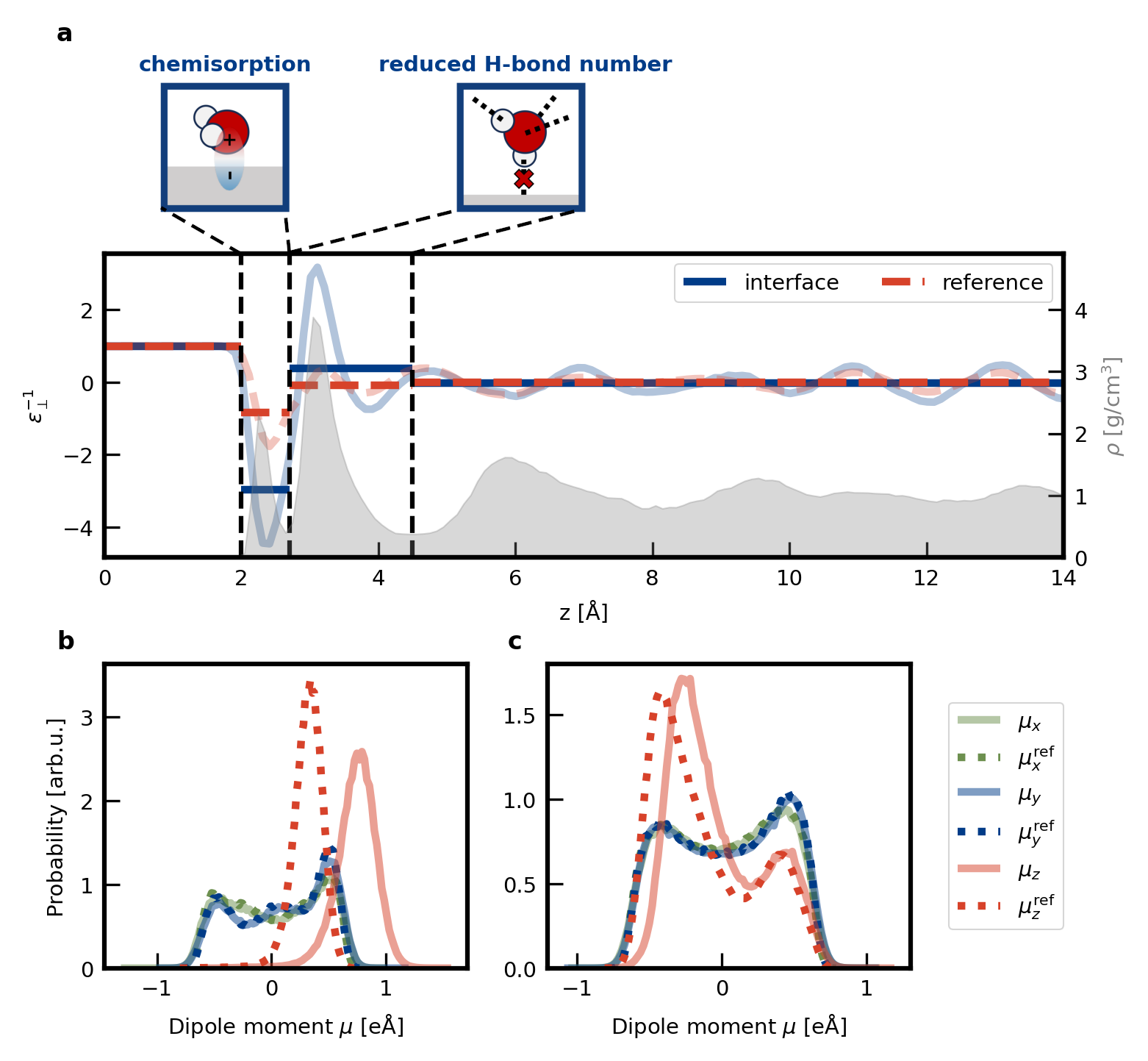}
  \caption{\textbf{The dielectric profile of water at the Pt/water interface.} \textbf{a} The light lines in the background represent the calculated dielectric profiles, including the effects of the intramolecular variations in the electric field. The dark lines in the foreground represent the regional average of the dielectric constant. While the blue lines refer to the calculation at the Pt/water interface, the red line refers to the dielectric profile calculated using the same trajectories but with the bulk water dipole moment (i.e., in the ``reference calculation''). The shaded area refers to the water density profile at the interface for the identification of the chemisorbed water layer and the physisorbed water layer. The two insets on the top illustrate the molecular origins of the electronic polarisation in the chemisorbed water and the physisorbed water. Distributions of water dipole moments of the chemisorbed water layer and the physisorbed water layer are shown in \textbf{b} and \textbf{c}, respectively. The subscripts denote the components in the x-, y-, or z-direction. The superscripts of ``ref" refer to the results from the reference calculation.}
  \label{fig:inveps}
\end{figure}

For non-interacting, homogeneous dielectrics, its susceptibility (i.e., $\varepsilon-1$) is expected to be proportional to densities (e.g., shaded area in Fig.~\ref{fig:inveps}a). However, water close to Pt(111) surfaces with higher densities than bulk water does not show a higher dielectric constant due to intricate metal-water interactions (see blue line in Fig.~\ref{fig:inveps}a). 
% At the Pt(111)/water interface, the chemisorbed water and the physisorbed water layer can be identified, which are expected to show distinct dielectric behaviours from the bulk water. Therefore, we calculate the average dielectric constant within these two layers, as shown in Fig.~\ref{fig:inveps}a. 
Particularly, a negative dielectric constant is observed at the region of 2-\SI{2.7}{\angstrom}, where the chemisorbed water is located. In contrast to the normal dielectric constant $\varepsilon$ (i.e., $0\le\varepsilon^{-1}\le1$), the negative dielectric constant indicates the over-screening of the electric field by the chemisorbed water layer\cite{tranNegativeDielectricConstant2023}. The more pronounced the over-screening effect, the more negative the value of the inverse dielectric constant $\varepsilon_\perp^{-1}$ in Eq.~\ref{eq:inveps}. The negative dielectric constant agrees with the negative capacitive response of the water chemisorption proposed in the previous work \cite{leMolecularOriginNegative2020}. In contrast, the physisorbed water layer shows a small average dielectric constant of $\sim$3, which is close to the dielectric constant used in the traditional EDL model \cite{conwayDielectricConstantSolution1951}.

To estimate the influence of the electronic polarisation in water induced by the interface, we calculate a dielectric profile with the water dipole moment set to the bulk value, denoted as the ``reference calculation'' in the following (see details in Supplementary Sec.4). Comparing the blue line and the red line in Fig.~\ref{fig:inveps}a, over-screening of the electric field in the chemisorbed water layer can be attributed to two reasons. 
On the one hand, as indicated by the negative values of the red line, over-screening exists due to water structuring at the interface.
On the other hand, the presence of chemisorption increases the water dipole moment and enhances the over-screening effect, based on the fact that the blue line becomes more negative.
% On the one hand, molecular reorientation is driven by the chemisorption rather than the electrostatic interaction, as a result of which the energy penalty from the over-screening is compensated by the water-metal bonding. This can be validated by the fact that the over-screening effect still exists in the red line, i.e., in the absence of electronic polarisation induced by the interface. 
% On the other hand, the presence of chemisorption increases the water dipole moment and enhances the over-screening effect.
The latter, i.e., how the electron polarization induced by chemisorption influences the dielectric constant, can be validated by the probability distribution of chemisorbed water dipole moments.
As shown in Fig.~\ref{fig:inveps}b, the molecular dipole moments of chemisorbed water increase in the z-direction (i.e., the direction perpendicular to the surface), while those in the x- and y-directions are almost identical after taking the electronic polarisation into account. Overall, the molecular origins of the dielectric property of the chemisorbed water layer shown by our calculation agree with the pictures proposed in previous studies\cite{leDeterminingPotentialsZero2017,leMolecularOriginNegative2020}.

In contrast to the chemisorbed water layer, the dielectric property of the physisorbed water layer is often believed to be mainly influenced by restricted molecular reorientation at the interface rather than electronic polarisation\cite{fumagalliAnomalouslyLowDielectric2018,leMolecularOriginNegative2020}. Interestingly, a distinct picture can be obtained from Fig.~\ref{fig:inveps}a. In the reference calculation, the physisorbed water layer exhibits an average dielectric constant comparable to the bulk region, which is significantly higher than the result with the effect of electronic polarisation. This indicates that the low dielectric constant observed in the physisorbed water layer at the interface is largely influenced by electronic polarisation rather than the widely accepted restricted reorientation of water molecules. A picture at the atomic level is revealed in Fig.~\ref{fig:inveps}c, in which the dipole moments of physisorbed water in the z-direction become smaller due to the influence of electronic polarisation. The reduced water dipole moment can be further rationalised as the results of the lower number of hydrogen bonds\cite{gregoryWaterDipoleMoment1997,kempInterpretationEnhancementWater2008} in the physisorbed water layer compared to the bulk water region. As shown by the inset in Fig.~\ref{fig:inveps}a, the physisorbed water with a reduced dipole points one hydrogen atom towards electrodes and loses one hydrogen-bond donor\cite{leStructureMetalwaterInterface2018}. 

With the help of the ec-MLP, we extend the timescale of MD simulations to \SI{1}{ns} and hence are able to calculate the dielectric profile at the Pt/water interface. In the chemisorbed water layer, an over-screening of the electric field is observed due to the synergy of the ordered molecular configurations and the electronic polarisation induced by chemisorption. Moreover, the average dielectric constant in the physisorbed water layer illustrates the significant influence of electronic polarisation in the absence of chemisorption, which, to the best of our knowledge, has rarely been discussed before. Overall, the dielectric profile calculated from the MLP sheds light on the effect of electronic polarisation on the dielectric response at the interface, which lays the foundation for future exploration of tuning interfacial dielectric properties and optimising the reaction activity in electrocatalysis.
% Additionally, the dielectric profile is instructive for the parameterisation of the implicit solvation models. 

\section{Discussion}

In this work, we introduce the ec-MLP model, which is feasible to describe the dielectric response at the electrode/electrolyte interfaces in a hybrid representation. The ec-MLP shows a high accuracy in predicting not only the water chemisorption but the potential-dependent interfacial water structures and differential capacitances. In particular, the Pt(111)/KF(aq) interface is chosen as the model system, which can only be simulated accurately when both the electronic structure and the molecular dynamics effect are taken into account. 
The success of ec-MLP in this challenging model system illustrates the feasibility of the method. 
Furthermore, we also calculate the dielectric profile at the Pt(111)/water interface and gain new insight into the effect of the electronic polarisation on the dielectric properties at the interface. While the electronic polarisation induced by chemisorption results in larger water dipole moments and over-screening of electric fields, that induced by fewer hydrogen bonds leads to smaller water dipole moments and lower dielectric constant.
Overall, the insight into the interfacial dielectric properties provided by the ec-MLP is inspiring for further studies of, for example, pH and ion effects, which are highly sensitive to the electronic structure at the interface. Additionally, we anticipate this method will be utilised to investigate some important processes with extended timescales at the electrochemical interface, e.g., electrocatalytic reactions \cite{chengFreeEnergyBarriersReaction2015,choiHighlyActiveStable2020}, formation of solid–electrolyte interphases (SEI) \cite{wangReviewModelingAnode2018,qinComputationalInvestigationLiF2024}, and electrode corrosion \cite{surendralalFirstPrinciplesApproachModel2018,arulmozhiNanoscaleMorphologicalEvolution2020}. 

Aiming for a wider application of the method, it is worthwhile to discuss how the MLP works under different conditions in a general way. At the electrochemical interfaces, there usually exist (electro-)chemical reactions, the treatment of which could be different in the MLP. For example, when the chemisorption happens at the interfaces, the electrons of the adsorbate polarise \cite{schmicklerInterfacialElectrochemistry2010}, which can be described by the displacements of the WCs in the MLP. As illustrated in Supplementary Sec.5, the (partial) charge transfer during the chemisorption can be captured via the electron redistribution within the adsorbates under constant potential constraint. In contrast, dealing with the electron transfer in the electrolyte is more complicated. Currently, the charge of the WCs and hence the oxidation states of the elements is fixed in the ec-MLP. Aiming for the free energy change of an electron transfer process, two MLPs with different oxidation states can be combined \cite{frenkelUnderstandingMolecularSimulation2002,wangAutomatedWorkflowComputation2022}. 

At the end of the discussion, we propose some possible strategies to improve the ec-MLP in the future. Currently, only the electrostatic interaction is minimised when solving the atomic charges in the Siepmann-Sprik model. In fact, we expect the chemical potential to also play a role, especially for the substrates with multiple elements. It is therefore promising to take advantage of the ML-based charge equilibrium model  \cite{dufilsPiNNwallHeterogeneousElectrode2023,shaoFinitefieldCouplingLearning2022,koAccurateFourthGenerationMachine2023}, which allows a higher generalisability of the ec-MLP. In addition, calculation of atomic forces of electrode atoms from the negative derivative of energy is computationally demanding. In this work, we approximate the forces of the electrode atoms as the sum of the electrostatic atomic forces and the short-range atomic forces from the derivative of the short-range energy, ignoring the dependency of the charges on the positions of the electrode atoms. The approximation can be justified as having a minor effect on this work based on the good performance of the ec-MLP. Nevertheless, more rigorous calculation of the atomic forces of the electrode atoms might be necessary when investigating, for example, the dynamic structure evolution of the electrode during the electrochemical processes \cite{grosseDynamicTransformationCubic2021}. In order to overcome the limitation, the automatic differentiation framework can be integrated into the ML model in future development\cite{wangDMFFOpenSourceAutomatic2023}.

% The capacitance values are slightly lower than the experimental results at high surface densities (i.e., 10 $\mu$F/cm$^2$ versus 20 $\mu$F/cm$^2$). The discrepancy might be attributed to the omission of the dielectric response at the high-frequency limit (i.e., electronic dielectric response) in the MLP. Nevertheless, this discrepancy can be approximately corrected by adding a constant value to the capacitance curve, considering the little electric field dependency of the electronic dielectric response \cite{schwarzElectrochemicalInterfaceFirstprinciples2020}. 

\section{Methods}

\subsection{DFT calculations for the dataset used in the machine learning model training}

All the density functional theory calculations were performed using CP2K/Quickstep package \cite{kuhneCP2KElectronicStructure2020}. In CP2K/Quickstep, a mixed Gaussian and plane waves approach \cite{vandevondeleQuickstepFastAccurate2005} is implemented for DFT calculations, in which the orbitals are described by an atom-centered Gaussian-type basis set and an auxiliary plane wave basis set is used to represent the electron density in the reciprocal space. The 2s and 2p electrons of O; 3s, 3p, and 4s electrons of K; 2s and 2p electrons of F; and 5d and 6s electrons of Pt are treated as valence, and the rest core electrons were represented by Goedecker-Teter-Hutter pseudopotentials \cite{goedeckerSeparableDualspaceGaussian1996,hartwigsenRelativisticSeparableDualspace1998}. The Gaussian basis sets were double zeta with one set of polarisation functions. The plane wave energy cutoff was set to 800 Ry. The Perdew-Burke-Ernzerhof (PBE) functional \cite{perdewGeneralizedGradientApproximation1996} was used to describe the exchange-correlation effects, with the Grimme D3 correction \cite{grimmeConsistentAccurateInitio2010} accounting for the dispersion interactions. The MLWCs were calculated from the DFT calculation by localising the occupied orbitals with a convergence threshold 0.01. The MLWCs were then associated with certain central atoms if the distances between the MLWCs and the atoms are smaller than \SI{1}{\angstrom}. 

\subsection{Training machine learning models for Wannier centroids}

% In the insulating systems, the maximally localised Wannier centres (MLWCs) can be calculated by localising the Bloch orbitals.
% According to the modern theory of polarisation, these Wannier centres can be used to calculate the polarisation of the system, which is necessary for the calculation under finite electric fields. [cite: modern theory of polarisation]
% In the absence of the electron transfer, the MLWCs can be assigned to the specific atoms and the centroids of the MLWCs can be calculated as the average of the MLWCs assigned to the same atom.
% Accroding to the previous work, the positions of the Wannier centorids only depend on their local environment, which can be accurately predicted by the machine learning models for atomic tensorial properties with local descriptors.  

In this work, the Deep Wannier (DW) model \cite{zhangDeepNeuralNetwork2020} was used to predict the positions of the WCs relative to the associated nuclei based on the local chemical environment. In this model, the nuclei carry the charges of the number of their valance electrons, i.e., \SI{6}{e}, \SI{1}{e}, \SI{9}{e}, \SI{7}{e}, and \SI{0}{e} for O, H, K, F and Pt atoms, respectively. The WCs with a charge of \SI{-8}{e} were associated with the O, K, and F atoms. When training the DW model, the descriptor was chosen as the Deep Potential Smooth Edition (DeepPot-SE) descriptor \cite{zhangEndtoendSymmetryPreserving2018} with a cutoff of \SI{5.5}{\angstrom}. The number of neurons in each hidden layer of the embedding net and the fitting net were [25, 50, 100] and [240, 240, 240], respectively. 
The DW model was trained with $5\times 10^5$ batches with a batch size of 1. The learning rate was set as starting from $1\times 10^{-3}$ and decaying to $1\times 10^{-8}$ with a step width of 5000 and rate of 0.95. 

\subsection{Training machine learning models for potential energy surfaces}

The training codes for total energies were implemented based on the Deep Potential Long Range (DPLR) method \cite{zhangDeepPotentialModel2022} in the software DeepMD-kit \cite{zengDeePMDkitV2Software2023}. As mentioned above, total energies can be decomposed into the short-range part and the long-range electrostatic part. The electrostatic contributions in total energies and atomic forces were calculated from the charge distributions based on the Particle-Particle Particle-Mesh (PPPM) algorithm \cite{dardenParticleMeshEwald1993}. The charge distributions of the electrolyte were predicted using the DW model, while those of the electrode were predicted using the Siepmann–Sprik model. The atomic charges of the electrode in the Siepmann–Sprik model follow Gaussian distributions centring at the atom sites. The Gaussian spread of \SI{1.6}{\angstrom} was chosen as the atomic radius of Pt used in the original charge equilibrium method \cite{rappeChargeEquilibrationMolecular1991}. 
% Notably, as shown in Supplementary Sec.4., the accuracy of the machine learning potential is insensitive to the value of the Gaussian width. 
The remaining short-range contributions can be trained via a neural network. When training the potential, the descriptor was chosen as the Deep Potential Smooth Edition (DeepPot-SE) descriptor \cite{zhangEndtoendSymmetryPreserving2018} with a cutoff of \SI{5.5}{\angstrom}. 
The number of neurons in each hidden layer of the embedding net and the fitting net were [25, 50, 100] and [240, 240, 240], respectively. 
The machine learning potential was trained with $5\times 10^5$ batches with a batch size of 1. 
The learning rate was set as starting from $1\times 10^{-3}$ and decaying to $1\times 10^{-8}$ with a step width of 5000 and rate of 0.95. 
The prefactor for the energy in the loss function started from 0.001 and ends at 1, while that for the atomic forces started from 1000 and ends at 1.

\subsection{Molecular dynamics simulation and calculation of interfacial dielectric properties}

The MD simulations in this work were performed with the software LAMMPS \cite{thompsonLAMMPSFlexibleSimulation2022}. The \verb|USER_DEEPMD| \cite{zengDeePMDkitV2Software2023} and \verb|ELECTRODE| \cite{ahrens-iwersELECTRODEElectrochemistryPackage2022} packages were modified to combine the DW model and the Siepmann–Sprik model. 

The MD simulations were performed in the canonical ensemble (NVT) using a timestep of 0.5 fs. In all simulations, the temperature was set as \SI{330}{K} with the damping time of \SI{50}{fs} for the Nosé-Hoover thermostat. The Ewald screening parameter was set as \SI{0.4}{\angstrom^{-1}} and the grid width in each dimension for the PPPM algorithm \cite{dardenParticleMeshEwald1993} was set as \SI{0.5}{\angstrom^{-1}}, which keeps consistent with the setup used in training the ec-MLP. 

The Pt(111)/electrolyte interfaces used in this work were described in a $(16.869\times14.609\times41.478)$ \si{\angstrom^3} periodic supercell model with a (6$\times$6) 6-atomic-layer Pt slab. The lattice constant of the Pt was \SI{3.98}{\angstrom}, which is consistent with the previous AIMD simulations \cite{leDeterminingPotentialsZero2017, leStructureMetalwaterInterface2018, leMolecularOriginNegative2020}. The electrolyte was filled between the space of \SI{30}{\angstrom} between two Pt surfaces. For all simulations, the water density in the bulk region (i.e., $\le$\SI{9}{\angstrom} from both surfaces) is (1$\pm$0.05) g cm$^{-3}$. All Pt atoms in the electrode were frozen during the simulations, which is required by the algorithm used in the \verb|ELECTRODE| package.

The constant charge method was used to investigate the potential-dependent properties at the interfaces, as shown in Fig.~\ref{fig:validation_simulation}. For each potential data point corresponding to a certain surface charge density, five simulations were performed with independent initial configurations. In each simulation, the system was pre-equilibrated for \SI{20}{ps} and sampled for \SI{100}{ps}.
% , which is slightly longer than the setup used in the previous AIMD simulation \cite{leMolecularOriginNegative2020}. 
% The standard deviations of the calculated potentials at a given surface charge density from the simulations are shown as the blue shadow in Fig.~\ref{fig:validation_simulation}c. 

The constant potential method was used to calculate the dielectric profile at the Pt/water interface. The simulations with the macroscopic electric fields of 0.0 V/\si{\angstrom} and 0.024 V/\si{\angstrom} were performed for \SI{1}{ns}. 
The convergence tests for the dielectric profile calculation are shown in Supplementary Sec.6.

\begin{acknowledgement}

  J.-X.Z. gratefully acknowledges Xiamen University and \textit{i}ChEM for a Ph.D. studentship. J.C. gratefully acknowledges funding from the National Science Fund for Distinguished Young Scholars (Grant No. 22225302), the National Natural Science Foundation of China (Grant Nos. 92161113, 21991151, 21991150, and 22021001) and the Fundamental Research Funds for the Central Universities (Grant Nos. 20720220008, 20720220009, 20720220010), Laboratory of AI for Electrochemistry (AI4EC), and IKKEM (Grant Nos. RD2023100101 and RD2022070501). J.-X.Z. also thanks Dr.~Jia-Bo Le, Dr.Xiao-Hui Yang, Dr.~Katharina Doblhoff-Dier and Dr.~Jun Huang for helpful discussions.

\end{acknowledgement}

% \begin{suppinfo}

% This will usually read something like: ``Experimental procedures and
% characterization data for all new compounds. The class will
% automatically add a sentence pointing to the information on-line:

% \end{suppinfo}

\section{Author contributions}

% J.-X.Z. conceived this work, designed the project, implemented the method and performed the simulation. J.-X.Z. and J.C. analysed the results and wrote the manuscript.
J.C. and J.-X.Z. conceived this work and designed the project. J.-X.Z. implemented the method and performed the simulation. J.-X.Z. and J.C. analysed the results and wrote the manuscript.

\section{Correspondence}
Correspondence to Jia-Xin Zhu or Jun Cheng.

\section{Competing interests}
The authors declare no competing interests.

\section{Data availability}
The data that support the findings of this study are available from the corresponding authors upon reasonable request.

%%%%%%%%%%%%%%%%%%%%%%%%%%%%%%%%%%%%%%%%%%%%%%%%%%%%%%%%%%%%%%%%%%%%%
%% The appropriate \bibliography command should be placed here.
%% Notice that the class file automatically sets \bibliographystyle
%% and also names the section correctly.
%%%%%%%%%%%%%%%%%%%%%%%%%%%%%%%%%%%%%%%%%%%%%%%%%%%%%%%%%%%%%%%%%%%%%
\bibliography{reference}

\end{document}